\documentclass{ws-p10x7}

\newcommand{\ra}{\rightarrow}
\newcommand{\bbb}{b\bar{b}}

\begin{document}

\title{Open Charm and Beauty Production at HERA}

\author{Felix Sefkow} 

\address{(on behalf of the H1 and ZEUS Collaborations) \\ 
        Physik-Institut der Universit\"at Z\"urich,
        Winterthurerstr.\ 190, CH-8057 Z\"urich,
Switzerland \\ E-mail: felix.sefkow@desy.de}

\twocolumn[\maketitle\abstract{
Selected new results from the H1 and ZEUS collaborations 
on $ep$ interactions at 300 - 318 GeV centre-of-mass energy 
are presented.
The full pre-upgrade integrated luminosity of HERA of 110 pb$^{-1}$ is used.
Charm cross sections are measured 
up to high values of $x_B$ and $Q^2$
and are found to be well described 
by NLO QCD in the 3 flavour scheme.
Orbitally excited $D$ mesons are observed; 
radial excitations are searched for, but are not seen.
The first $b$ cross section measurement  
is confirmed with a lifetime based method, 
establishing the excess over NLO QCD.
}]

\section{Charm}

Thanks to the excellent HERA performance the available statistics 
has strongly increased. 
ZEUS now has a signal of 27,000 $D^{\ast}$ decays 
in the ``golden'' mode $D^{\ast}\ra D^0\pi^+\ra K^-\pi^+\pi^+$.
This wealth of data (similarly at hand for H1) allows  
perturbative QCD to be tested with charm production data
in an extended kinematic range 
and opens the possibility for charm spectroscopy at HERA. 

In QCD, heavy quark production in $ep$ interactions predominantly
proceeds via boson gluon fusion (BGF), where a quark anti-quark pair is 
created in the interaction of a photon 
with a gluon in the proton (3~Flavour scheme). 
At four-momentum transfers much higher than the charm quark mass, 
$Q^2 \gg m_c^2$, 
such a description becomes inaccurate, 
and a treatment in terms of charm densities in the proton 
may be more adequate. 

\begin{figure}
\epsfxsize165pt
\figurebox{}{}{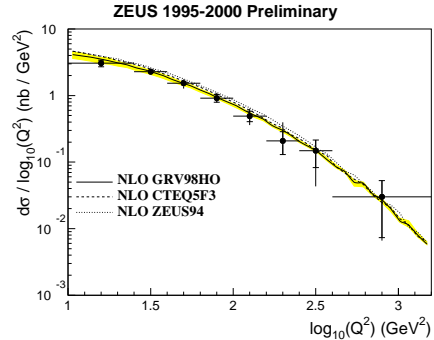}
%
\caption{$D^{\ast}$ cross section in DIS vs. NLO QCD
in the 3~Flavour scheme (shaded: $m_c =  1.3 - 1.6\,$GeV).}
\label{fig:dstdis}
\end{figure}

The single-differential $D^{\ast}$ cross sections 
measured in deep inelastic scattering (DIS) by ZEUS~\cite{zeusdis}  
as a function of Bjorken-$x$ and $Q^2$ 
now cover a range up to $x_B\simeq 0.1$ and $Q^2\simeq 1000\,$GeV$^2$.
(Fig.~\ref{fig:dstdis})
They are compared 
with NLO QCD calculations in the 3~Flavour scheme~\cite{hvqdis},
which use as input 
gluon densities from global fits~\cite{globfit}  
or a parameterization extracted
from scaling violations of the proton structure function $F_2$,
measured at HERA. 
Good agreement is seen, 
showing that the BGF picture provides an overall consistent description
of charm production and inclusive DIS up to high $x_B$ and $Q^2$. 
 
The spectrum of non-strange $D$ mesons is only partially established 
experimentally. 
Apart from the lowest mass $D$ and $D^{\ast}$ states, 
the narrow excited P-wave mesons $D_1(2420)$ and $D_2^{\ast}(2460)$
have been firmly identified, 
with spin-parity $J^P=1^+$ and $2^+$. 
A narrow state interpreted as radially excited $D^{\ast\prime\pm}$ has been 
observed by DELPHI~\cite{delphidstp}, 
but was not confirmed by OPAL 
and CLEO searches~\cite{otherdstp}.

\begin{figure}[htb]
\epsfxsize220pt
\figurebox{}{}{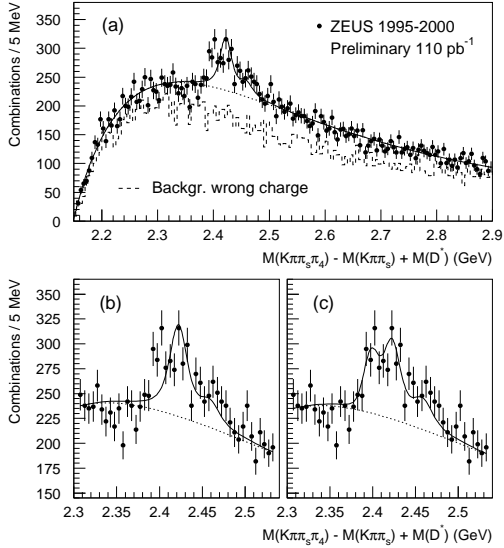}
\caption{Mass difference distribution for $D^{\ast\ast 0}$ candidates.
The curves show fits using mass and helicity angle information.}
\label{fig:dstp}
\end{figure}

ZEUS report~\cite{zeusspect} the observation of orbitally excited $D^0_1$ and 
$D^{\ast 0}_2$ mesons in the decay channel 
$D_J^{(\ast)0}\ra D^{\ast +}\pi^-$ + c.c.  
From a fit to the invariant mass
(Fig.~\ref{fig:dstp})
and $\pi^-$ helicity angle distributions
they extract relative production rates of 
\begin{eqnarray}
\frac{D_1^0\ra D^{\ast +}\pi^-}{D^{\ast +}} & = & 
3.40 \pm 0.42\; ^{+0.78}_{-0.63}\; \%
\nonumber \\[4pt]
\frac{D_2^{\ast 0}\ra D^{\ast+}\pi^-}{D^{\ast +}} & = & 
1.37 \pm 0.40\; ^{+0.96}_{-0.33} \; \%  \nonumber
\end{eqnarray}
A narrow enhancement ($\sim 4\sigma$) is seen at $m_{D^{\ast}\pi}=$
2398 MeV
and included in the fit (Fig.~\ref{fig:dstp}c),
but no definite interpretation of this signal is given yet. 

\begin{figure}[htb]
\epsfxsize220pt
\figurebox{}{}{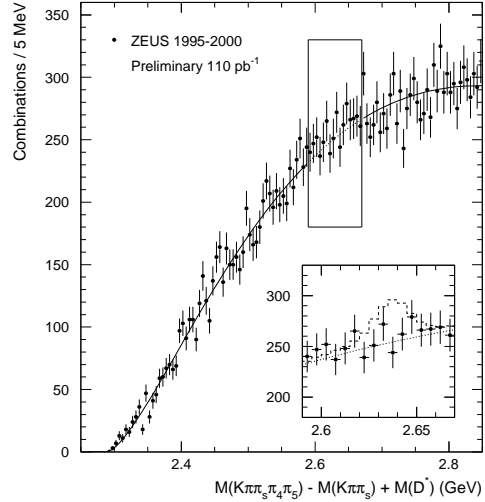}
\caption{Mass difference distribution for $D^{\ast\prime\pm}$ candidates.
The search region contains $91\pm75$ candidates over the fitted background. 
The insert shows a Monte Carlo signal normalized to the quoted limit.}  
\label{fig:dstpp}
\end{figure}

Radially excited states are searched for in the channel 
$D^{\ast\prime\pm}\ra D^{\ast +}\pi^+\pi^-$ + c.c.
No signal is seen (Fig.~\ref{fig:dstpp}), 
so that an upper limit is quoted:
\begin{eqnarray}
\frac{D^{*\prime +}\ra D^{*+}\pi^+\pi^-}{D^{*+}} < 2.3\,\%\;\;
\mbox{\rm (at 95\% C.L.)} \nonumber
\end{eqnarray}
which indicates that the search has a sensitivity corresponding
to about the size of the claimed DELPHI signal. 
Since at HERA almost all charmed mesons originate 
from prompt charm production, and feed-down from beauty can be neglected,
a rather tight limit on $D^{\ast\prime\pm}$ production in charm fragmentation 
can be set (at 95\% C.L.):
\begin{eqnarray}
f(c\ra D^{*\prime +}) \cdot {\rm BR} (D^{*\prime +}\ra D^{*+}\pi^+\pi^-)
< 0.7 \%\nonumber
\end{eqnarray}

\section{Beauty}

\begin{figure}[htb]
\epsfxsize170pt
\figurebox{}{}{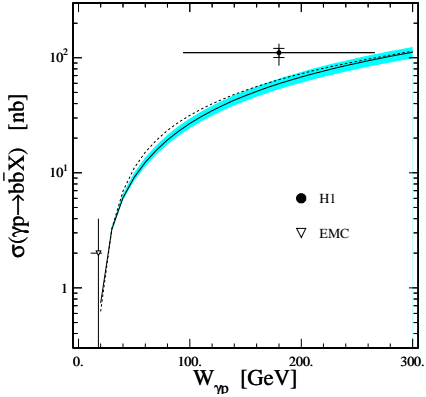}
\caption{$b$ photoproduction cross section vs.\ NLO QCD,
using different proton structure functions (shaded: scale uncertainty).}
\label{fig:h1bnlo}
\end{figure}

Beauty production at HERA is suppressed with respect to charm by two 
orders of magnitude. 
The measurements so far rely on inclusive semi-leptonic decays,
using as signature the high mass of the $b$ quark by observing the 
transverse momentum $p_T^{rel}$ of the lepton relative to a jet, 
and also its long lifetime by observing 
tracks from secondary vertices.  
The first measurement by H1~\cite{h1b}, using the $p_T^{rel}$ method, 
revealed a $b$ photoproduction cross section almost a factor of 2 above 
theoretical prediction~\cite{frixi} 
(Fig.~\ref{fig:h1bnlo}).

\begin{figure}[htb]
\epsfxsize190pt
\figurebox{}{}{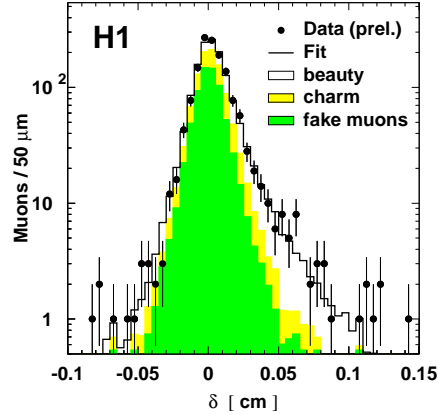}
\caption{Signed impact parameter distribution for muons.
(The sign depends on whether the track intersects the jet axis upstream or 
downstream (+) of the primary event vertex.)}
\label{fig:delta}
\end{figure}

\begin{figure}[htb]
\epsfxsize160pt
\figurebox{}{}{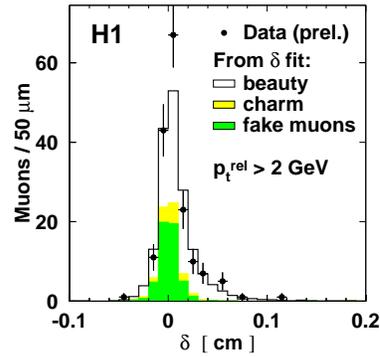}
\caption{Impact parameter distribution for muons with $p_T^{rel}>2\,$GeV.
with the absolute prediction from the fit to the full sample.}
\label{fig:deltapure}
\end{figure}

The new H1 measurement~\cite{h1blto} also uses photoproduction 
dijet events, where now at least one muon
is measured in the two-layer silicon vertex detector. 
The signed impact parameter $\delta$ is determined 
in the plane transverse to the beam, axis, 
and the distribution is decomposed by a maximum likelihood fit
which adjusts the relative contributions from beauty, charm and 
fake muons to the sample
(Fig.~\ref{fig:delta}).
The fit describes the data well and 
translates into a $b$ cross section
that, using an independent 
signature and new data,
confirms the published result, based on 
1996 data and a different set of cuts. 
The $\delta$ spectrum for a sample with higher $b$ purity, 
obtained by a cut 
$p_T^{rel}>2\,$GeV 
(Fig.~\ref{fig:deltapure})
agrees well with the 
prediction from the $\delta$ fit to the full
sample. 
Since the two observables are consistent, 
they can be combined    
in a likelihood 
fit of the two-dimensional $(\delta, p_T^{rel})$ distribution.
The result, averaged with the published number, is
\begin{eqnarray}
\sigma (ep\ra\bbb X\ra\mu X) =
(170\,\pm\,25)\,{\rm pb} \nonumber
\end{eqnarray}
in the range 
$Q^2<1$ GeV$^2$, 
$0.1<y<0.8\,$, 
$p_T(\mu)>2$GeV,
$35^\circ<\theta_(\mu)<130^\circ$.
This is higher than the NLO QCD prediction 
of (104 $\pm$ 17)~pb based on~\cite{frixi}.
Such a discrepancy between experiment and NLO QCD
is now established in both $ep$ and 
$\bar{p}p$ interactions.

\end{document}